\documentclass[12pt]{iopart}

\usepackage{amsfonts,epsf}

\newcommand{\Even}{{even}}
\newcommand{\Odd}{{odd}}

\def\ggg{\eta}

\newcommand{\blank}[1]{}
\newcommand{\cF}{{\ensuremath{\cal F}}}
\newcommand{\cG}{{\ensuremath{\cal G}}}
\newcommand{\cH}{{\ensuremath{\cal H}}}

\def\cbI#1#2#3#4#5#6#7#8{
\setlength{\unitlength}{#1sp}%
\centering{
\begin{picture}(1500,757)(4500,-2483)
\thicklines
{\put(4850,-1860){\line( 1,0){800}}}%
{\put(4850,-2460){\line( 1,0){800}}}%
{\put(5250,-1860){\line( 0,-1){600}}}%
\put(4775,-2460){\makebox(0,0)[rc]{$#2$}}
\put(4775,-1860){\makebox(0,0)[rc]{$#3$}}
\put(5750,-1860){\makebox(0,0)[lc]{$#4$}}
\put(5750,-2460){\makebox(0,0)[lc]{$#5$}}
\put(5300,-2160){\makebox(0,0)[lc]{$#6$}}
\end{picture}}}%

\def\cbb#1#2#3#4#5#6#7#8{
\setlength{\unitlength}{#1sp}%
\begin{picture}(2800,1400)(3901,-2683)
\thicklines
{\put(4800,-1860){\line( 0,-1){600}}}%
{\put(5700,-1860){\line( 0,-1){600}}}%
{\put(4200,-2460){\line( 1, 0){2100}}}%
\put(4150,-2460){\makebox(0,0)[rc]{$#2$}}
\put(4800,-1800){\makebox(0,0)[cb]{$#3$}}
\put(4800,-2560){\makebox(0,0)[ct]{$#7$}}
\put(5250,-2380){\makebox(0,0)[cb]{$#4$}}
\put(5700,-1800){\makebox(0,0)[cb]{$#5$}}
\put(5700,-2560){\makebox(0,0)[ct]{$#8$}}
\put(6400,-2460){\makebox(0,0)[lc]{$#6$}}
\end{picture}}%

\newcommand\Zb            {\mathbb{Z}}

\def\d{{\rm d}}
\def\cev#1{\langle #1 \vert}
\def\D#1{ \frac{\d#1}{2\pi i}}
\def\vec#1{\vert #1 \rangle}
\def\tfrac#1#2{{\textstyle{\frac{#1}{#2}}}}

\begin{document}

\rightline{kcl-mth-08-09}
\rightline{16 December 2008}
\title{Differential equations from null vectors of the Ramond algebra}

\author{P Giokas and G M T Watts}
\address{
Dept.\ of Mathematics, King's College London,
Strand, London WC2R 2LS -- UK
}

\ead{philip.giokas@kcl.ac.uk, gerard.watts@kcl.ac.uk}

\begin{abstract}
We consider chiral blocks of four
Ramond fields of the $N{=}1$ super Virasoro algebra where one of the
fields is in the (1,2) representation. 
We show how the null vector in the (1,2) representation determines the
chiral blocks as series expansions. We
then turn to the Ising model to find an algebraic method to determine
differential equations for the blocks of four spin fields. Extending
these ideas to the super Virasoro case, we find a first order
differential equation for blocks of four Ramond fields. We are able to
find exact solutions in many 
cases. We compare our blocks with results known from other methods.
\end{abstract}

\maketitle

\section{Introduction}
\label{sec.intro}

Conformal field theories in two dimensions can be used to describe
string theory from the worldsheet perspective and statistical
systems at a second order phase transition. The study of these field
theories is often manageable because the infinite dimensional symmetry
algebras which exist can reduce the field content to a finite number
of representations. In such minimal cases, there are `null vectors'
which should decouple from all correlation functions. As a result the
non-zero correlation functions satisfy differential equations which
enable one to solve the theory completely.

Each symmetry algebra has its own series of minimal models for which
this procedure works. The simplest case is that of minimal models of
the Virasoro algebra. In \cite{BPZ}, Belavin et al showed how to
relate correlation functions of arbitrary fields in a Virasoro minimal
model to those of a particular class of fields, called primary fields,
and in turn showed how differential equations for four-point functions
of primary fields could be found from the singular vectors that are
present in these models.

The Virasoro algebra can be extended by extra generators to include
supersymmetry, and the simplest of these extensions is the $N{=}1$
superconformal algebra. Although the $N{=}1$ superconformal algebra
includes the Virasoro algebra, there is an infinite set of minimal
models of the $N{=}1$ superconformal algebra which only includes a few
minimal models of the Virasoro algebra, and the methods of \cite{BPZ}
need to be generalised to find the differential equations satisfied by
the correlation functions of primary fields of the $N{=}1$
superconformal algebra.

The first thing to note is that the superconformal algebra comes in
two forms, called the Neveu-Schwarz (NS) and Ramond (R) algebras, and
accordingly has two classes of representations and fields associated
to these representations.  The differential equations for correlations
functions of four NS--type fields were worked out in \cite{FQSh3,Qiu},
but the extension to correlation functions of four Ramond fields has
problems \cite{mss1,mss2} and the solution to these problems presented
here requires an extension of ideas in \cite{FQSh3,mss2}.
In this paper we perform this extension and find differential
equations for correlation functions of four R--type fields. 

The structure of the paper is as follows: we first introduce the NS
and R algebras and describe their representations and singular
vectors. We discuss their three point functions and describe the
chiral blocks and how the singular vectors allow them to be found
level-by-level.   
Next we describe the toy model of the Ising model and show how one can
obtain a differential equation for the four-spin field correlation
function, as in \cite{mss2}.
Next we apply these ideas to
the Ramond correlation functions. 
This lets us write down a matrix-differential equation for the
correlation functions. 
We check these by comparison with known results and present the exact solution
for a variety of cases.

\section{Representation theory}
\label{sec.rep}

The algebra of chiral superconformal transformations in the plane is
generated by two fields, $L(z) = \sum_m L_m z^{-m-2}$ and 
$G(z) = \sum_m G_m z^{-m-3/2}$. According to the choice of boundary
conditions on $G(z)$, one may choose the labels $m$ on $G_m$ to be
integer or half-integer; the algebra in these two sectors are called
the Ramond and Neveu Schwarz algebras respectively. In both sectors
the generators obey 
\begin{eqnarray}
{}~[L_m,L_n]  &=&
\frac c{12}	 m(m^2-1)\delta_{m+n,0} + (m-n)L_{m+n}  \;,\\
 \{G_m,G_n\} &=& \frac c3
		(m^2 - \frac 14) \delta_{m+n,0} + 2 L_{m+n} \;,\\
{}~[L_m,G_n]  &=& (\frac m2 - n)G_{m+n}\;.
\end{eqnarray}
It is usual to adjoin the operator $(-1)^F$ to the superconformal
algebra, where $F$ is the fermion number operator which anti-commutes
with $G_m$ and commutes with $L_m$. This allows one to consider states
of definite 
parity and is essential in many constructions.
In a superconformal field theory the physically relevant
representations of the superconformal algebra are irreducible highest
weight representations. These are graded by $L_0$ eigenvalue, or level.
See \cite{FQSh3,MRCa1} for more details.

\subsection{NS representations}

Highest weight representations of the Neveu-Schwarz algebra have a state
of least $L_0$ eigenvalue $\vec h$ such that
\begin{eqnarray}
L_m |h\rangle &=& h \delta_{m,0} |h\rangle,\; m\ge 0
\label{eq.hw.l} \;, \\
G_m |h\rangle &=& 0,\; m\ge 1/2
\label{eq.hw.g} \;.
\end{eqnarray}
If one considers the extended algebra then it is usual to take the
highest weight state bosonic ($(-1)^F \vec h = \vec h$) or fermionic
($(-1)^F\vec h = -\vec h$) in which case the addition of the fermion
number operator does not alter the representation theory.
It is useful to parametrise $h$ and $c$ as 
\begin{eqnarray}
&&c(t) = 15/2 - 3/t - 3t \;,
\label{eq.c.t}
\\
&&h_{p,q}(t) = (1-pq)/4 + (q^2 - 1)t/8 + (p^2-1)/(8t)  \;, 
\end{eqnarray}
since there is a singular vector of $L_0$ eigenvalue $h + pq/2$ 
whenever $h=h_{p,q}(t)$ and  $c=c(t)$ and $p,q\in\Zb$, $p+q \in 2\Zb$
(see refs.\ \cite{Kac2,FeFu1,FeFu2}). 
Note that the highest weight state can be either
bosonic or fermionic, for example, the NS vacuum state $\vec 0$ is usually
regarded as bosonic and so the NS highest weight state $G_{-1/2}\vec 0$
is fermionic.
The highest weight representation $\cH_h$ is spanned by states of the form
\begin{equation}
L_{i_1}\ldots L_{i_m} G_{j_1}\ldots G_{j_n}\vec h \;,
\label{eq.verma}
\end{equation}
and the level of such as state is $(\sum i_p + \sum j_q)$.
States with even numbers of modes of $G$ have integer level and states
with odd numbers of modes have half-integer level and hence 
$\cH_h = \cH^{even}_h + \cH^{odd}_h$ where 
 $\cH^{even}_h$ is spanned by integer level states and  $\cH^{odd}_h$
by half-integer level states.

\subsection{R representations}

Highest weight representations of the extended Ramond algebra
have, in general, 
a two-dimensional highest weight space spanned by states 
$\vec{\lambda^\pm}$ of definite fermion parity $\pm 1$
satisfying
\begin{equation}
\begin{array}{rcll}
    L_m \vec{\lambda^\pm} 
&=& h_\lambda \delta_{m,0} \vec{\lambda^\pm}
\;, & m\ge 0 \nonumber \;,
\\
    G_m \vec{\lambda^\pm} 
&=& 0
\;, & m> 0 \;,
\\
    G_0 \vec{\lambda^\pm} 
&=& b^\lambda_{\pm} \vec{\lambda^\mp}
\;, 
\end{array}
\label{eq.hw.g.r}
\end{equation}
where $h_\lambda = \lambda^2 + c/24$ and $b^\lambda_+ = y \lambda,
b^\lambda_- = \lambda/y$ and $y$ can be chosen freely; representations
with different $y$ are equivalent.
Such a representation
has two singular vectors at level $pq/2$ whenever $p,q\in \Zb$, $p-q$
an odd integer, $c=c(t)$ and 
$\lambda=\lambda_{p,q}(t)$,
\begin{equation}
\lambda_{p,q}(t) =
{{ p - qt }\over{ 2 \sqrt{ 2 t }}} \;.
\label{eq:lpq}
\end{equation}
We shall mostly be concerned with the constraints arising from the
vanishing of the singular vectors at level 1 in the representation
$(1,2)$ which take the form
\begin{equation}
(\, L_{-1} + \frac{2t}{1 - 2t} G_{-1}G_0\, ) \vec{\lambda_{1,2}^\pm} = 0
\;.
\label{eq.nv12}
\end{equation}

\section{Ramond chiral blocks}
\label{sec.chiralblocks}

We would like to find the chiral blocks of four
Ramond fields
\begin{equation}
F(z) = \cev{\lambda_\infty^\pm} \phi^\pm_{\lambda_1}(1) 
       \phi_{\lambda_z}^\pm(z) \vec{\lambda_1^\pm}
\end{equation}
where the intermediate states lie in a particular NS representation.
We can do this step-by-step by first calculating the operator product
expansions (opes)
\begin{equation}
 \phi^\epsilon_\lambda(z)\; \vec{\mu^{\epsilon'}}
\;,
\label{eq:opes}
\end{equation}
and then forming the chiral block by summing over intermediate states.
To do this we need to define the vertex
operators $\phi_\lambda^\pm(z)$ of fixed fermion parity and we also need to
specify whether the NS representation in the intermediate channel is
bosonic or fermionic. 

For the field operators, we shall use
the definition in \cite{FQSh3}:
\begin{equation}
      K_m^+(w) \phi^\pm_\lambda(w)
\;\pm\; 
    i \phi^\pm_\lambda(w) K_{m-1/2}(w)w^{1/2}
\;=\;
    a^\lambda_\pm w^m \phi^\mp_\lambda(w)
\;,
\label{eq:kppk}
\end{equation}
where the combinations $K_m(w)$ and $K^+_m(w)$ are 
\begin{eqnarray}
K_m^+(w) &=&
	\oint_{0, |z|>|w|}  \!\!\!\!\!\!\!\! \!\!\!\! G(z)z^m \sqrt{z-w} \D z
= G_m -\frac{w}{2} G_{m-1} -\frac{w^2}8 G_{m-2}+\ldots
\nonumber\\
K_m(w) &=&
	\oint_{0, |w|>|z|}  \!\!\!\!\!\!\!\! \!\!\!\!
        G(z)z^m\sqrt{\textstyle\frac{w-z}{w}}\D z
= G_m - \frac{1}{2w}G_{m+1} - \frac{1}{8w^2}G_{m+2}+\ldots
\label{eq:Ks}
\end{eqnarray}
$a_\lambda^+ = \lambda x$, $a_\lambda^- = \lambda/x$
and $m$ can be integral or half-integral as circumstances dictate
(here $a^\pm$ play the same role as $b^\pm$ in \eref{eq.hw.g.r} and
$x$ as $y$).  

It turns out that the four opes in \eref{eq:opes} depend on only two
quantities. It is convenient to form the following four linear
combinations:
\begin{eqnarray}
    \cF^\pm_{\lambda\mu}(z) 
&=& \phi_\lambda^+(z)\, \vec{\mu^+}\; \pm\; ixy
    \phi_\lambda^-(z)\, \vec{\mu^-}
\;,\;\;
\nonumber\\
    \cG^\pm_{\lambda\mu}(z) 
&=& x\phi_\lambda^-(z)\, \vec{\mu^+}\; \pm\; iy
    \phi_\lambda^+(z)\, \vec{\mu^-}
\;,\;\;
\end{eqnarray}
We shall first consider the case where the NS highest weight state is
bosonic and the opes depend on the two constants $f^\pm_{\lambda\mu} = \cev h
\cF^\pm_{\lambda\mu}(1)$. 
Using \eref{eq:Ks} and the standard result for the Virasoro algebra
(see \eref{eq.l.c4}), we can calculate the overlaps
of $\cF^\pm_{\lambda\mu}(z)$ and $\cG^\pm_{\lambda\mu}(z)$. We list
here the first few, taking $z=1$ for convenience:
\begin{equation}
\renewcommand{\arraystretch}{1.4}
\begin{array}{rcl}
    \cev h G_{1/2} \,\cG^\pm_{\lambda\mu}(1) 
&=& (\lambda \pm \mu) \, f^\pm_{\lambda\mu}
\;,\\
    \cev h L_{1}   \,\cF^\pm_{\lambda\mu}(1) 
&=& (h + h_\lambda - h_\mu) \, f^\pm_{\lambda\mu}
\;,\\
    \cev h G_{3/2} \,\cG^\pm_{\lambda\mu}(1) 
&=& \frac12{(3\lambda \pm \mu)} \, f^\pm_{\lambda\mu}
\;,\\
    \cev h L_{1}G_{1/2} \,\cG^\pm_{\lambda\mu}(1) 
&=& (\lambda\pm\mu)(h + \lambda^2 - \mu^2 + \tfrac 12) 
    \, f^\pm_{\lambda\mu}
\;.\\
\end{array}
\end{equation}
If we take the NS representation to be fermionic rather than bosonic,
then $\cF^\pm$ and $\cG^\mp$ just swap roles in these equations, so
without loss of generality we consider henceforth the intermediate
channel to be bosonic.
Given these combinations $\cF^\pm, \cG^\pm$, it is possible to
define the chiral blocks and calculate them to be
\begin{eqnarray}
  \cbb{2300}{\lambda_\infty}{\lambda_1,1}{h^{\Even}}{\lambda_z,z}{\lambda_0}{\epsilon\vphantom{{}'}}{\epsilon'}
&=& \overline{\cF^\epsilon_{\lambda_1\lambda_\infty}\!(1)}
          \,  \cF^{\epsilon'}_{\lambda_z\lambda_0}(z)
\label{eq:ff} 
\\
&=& z^{h - h_z - h_0} 
  \left\{
1 + 
z \tfrac{(h {+} h_z {-} h_0)(h {+} h_1 {-} h_\infty)}{2h} +
  \ldots 
\;\right\}
\;,
\nonumber\\
  \cbb{2300}{\lambda_\infty}{\lambda_1,1}{h^{\Odd}}{\lambda_z,z}{\lambda_0}{\epsilon\vphantom{{}'}}{\epsilon'}
\quad
&=& \overline{\cG^\epsilon_{\lambda_1\lambda_\infty}\!(1)}
          \,  \cG^{\epsilon'}_{\lambda_z\lambda_0}(z) 
\label{eq:GG}\\
&=& z^{h - h_z - h_0+1/2} 
 \left\{ \;
\frac{(\lambda_z + \epsilon' \lambda_0)
      (\lambda_1 + \epsilon  \lambda_\infty)}{2h} + \ldots \;
 \right\}\;,
\nonumber \end{eqnarray}
where `even' denotes projection of the intermediate states onto
 $\cH_h^{even}$ and 
`odd', projection onto
 $\cH_h^{odd}$, where we again note that we have assumed that $\vec h$
 is bosonic.

\section{Singular vector decoupling}
\label{sec.field}

The vanishing of the state \eref{eq.nv12} imposes constraints on the
allowed fusions of the fields $\phi^\pm_{1,2}$ and in fact allows one
to determine the operator products $\cF^\pm_{\lambda\lambda_{1,2}}(z)$
and  $\cG^\pm_{\lambda\lambda_{1,2}}(z)$ recursively. 
We take 
$\lambda=\lambda_{1,s}$ as this simplifies the constraints on $h$, and
we shall denote $\cF^\pm_{\lambda_{1,s}\lambda_{1,2}}(1)$ by $\cF^\pm$
and $\cG^\pm_{\lambda_{1,s}\lambda_{1,2}}(1)$ by $\cG^\pm$ for simplicity.

If we use the equation (valid for any Virasoro primary field of weight
$h$)
\begin{equation}
 \phi_h(1) (L_{-1} - L_0) = (L_{-1} - L_0 +h)\phi_h(1)
\;,
\end{equation} and the relations \eref{eq:kppk}, we find that 
 $\cF^\pm$ and $\cG^\pm$ satisfy
\begin{eqnarray}
    (h_{1,s\pm 1} - L_0 + L_{-1}) \,\cF^\pm
&=& 
\mp \sqrt{\frac t2}K^+_{-1/2}     \,\cG^\pm
\;,
\\
    (h_{1,s\mp 1} - L_0 + L_{-1}) \,\cG^\pm
&=& 
\pm \sqrt{\frac t2}K^+_{-1/2} \,\cF^\pm
\;,
\end{eqnarray}
where we assume the representation $h$ to be bosonic. Taking $h$
fermionic simply swaps $\cF^\pm$ with $\cG^\pm$.
Considering the contribution of the highest weight state $\vec h$ to
$\cF^\pm$, we see that
\begin{equation}
    (h_{1,s\pm 1} - h) \,\cev h \cF^\pm
= 0 
\;,
\end{equation}
so that $\cF^\pm$ is only non-zero if $h=h_{1,s\pm 1}$. The two
allowed fusions and the operator products are thus
\begin{eqnarray}
&&\hbox{ $+$ channel:}\qquad  (1,2) \times (1,s) \longrightarrow (1,s{+}1)
\label{eq:+channel}
\\
& \cF^+ &= \vec h + \tfrac{s-1}s L_{-1}\vec h + \ldots
\nonumber\\
& \cG^+ &= -\tfrac 1s \sqrt{\tfrac 2t}G_{-1/2}\vec h 
+ \sqrt{\tfrac t2}\tfrac 1{st{+}2}\left(
 G_{-3/2} + \tfrac{2(t{-}2{+}st)}{st}L_{-1}G_{-1/2}\right)\vec h + \ldots
\nonumber\\[6mm]
&&\hbox{ $-$ channel:}\qquad   (1,2) \times (1,s) \longrightarrow (1,s{-}1)
\label{eq:-channel}
\\
& \cF^- &= \vec h + \tfrac{st{+}t-2}{st{-}2} L_{-1}\vec h + \ldots
\nonumber\\
& \cG^- &= -\tfrac{\sqrt{2t}}{2t{-}2} G_{-1/2}\vec h 
+ \sqrt{\tfrac t2}\tfrac 1{st{-}4}\left(
 G_{-3/2} - \tfrac{2(st{+}t{-}4)}{st{-}2}L_{-1}G_{-1/2}\right)\vec h + \ldots
\nonumber
\end{eqnarray}
Using these expressions it is then possible in principle to calculate
the chiral blocks 
order by order and turn the recursion relations for the operator
product expansions into differential equations on the chiral blocks.
Rather than carry this out in detail we will instead turn to a method
to derive differential equations directly.

\subsection{Differential equations from Singular vector decoupling}

We would like to find differential equations for a chiral block of the
form
\begin{equation}
F(z) = \cev{\phi_{\infty}} \phi_1(1) \phi_z(z) \vec{\lambda_{12}}
\;,
\label{eq.cf1}
\end{equation}
coming from the vanishing of \eref{eq.nv12},
where $\phi_\infty$, $\phi_1$ and $\phi_z$ are possibly matrix-valued
Ramond fields.
As a first step, we consider the contribution coming from the mode
$L_{-1}$. This is easy to calculate  using the relations for a
Virasoro primary field of weight $h$ 
\begin{eqnarray}
{}[L_m,\phi(z)] &=&
 z^{m+1}\phi'(z) + z^m h(m+1)\phi(z) 
\;.
\label{eq.l.c4}
\end{eqnarray}
The only complication is the need to remove the term in $\phi'_1(1)$
that would normally arise. This can be done by considering
\begin{equation}
L_{-1} \vec{\lambda_{12}} = (L_{-1} - L_{0} + h_{1,2})\vec{\lambda_{12}}
\end{equation}
to find
\begin{equation}
\cev{\phi_\infty} \phi_1(1) \phi_z(z) L_{-1} \vec{\lambda_{12}}
= \left[
  (z-1)\frac{\partial}{\partial z} + h_1 + h_z + h_{1,2} - h_\infty
  \right] F(z)
\;.
\label{eq:l-1}
\end{equation}
The difficulty now is to treat $G_{-1}$. We shall adapt some ideas
used in the construction of the co-invariant spaces that are used to
classify the fusion algebra of Virasoro and superconformal algebra
representations.

In this method, one finds linear combinations of modes which
annihilate the state $\cev{\phi_\infty}\phi_1(1)\phi_z(z)$. 
The simplest way to find these is to find polynomials $p(w)$ such that
$\cev{\phi_\infty}\phi_1(1)\phi_z(z) p(w)T(w)$ has at most a simple
pole at $1$, $z$ and $\infty$. The modes of $p(w)T(w)$ then take
simple values when acting on  $\cev{\phi_\infty}\phi_1(1)\phi_z(z)$. 
For example, for $m<0$
\begin{eqnarray}
 l_m &=& \oint_0 T(w) (1-w)(z-w)w^{m+1} \D w
     \\
&=& z L_m - (1+z)L_{m+1} + L_{m+2}
\end{eqnarray}
satisfies
\begin{equation}
\cev{h_\infty}\phi_1(1) \phi_z(z) l_m
= (1-z)(z^{m+1}h_z - h_1) \cev{h_\infty}\phi_1(1) \phi_z(z)
\;.
\end{equation}

As a toy example, we shall first apply this idea to the Ising model to
reproduce the result of \cite{mss2} before applying it to the
superconformal algebra and Ramond fields.

\section{The Ising model}

The Ising model can be formulated as the theory of a single free
fermion field $\psi(z)$. As with the superconformal algebra, this field
can have half-integer (NS) or integer (R) modes
\begin{equation}
 \psi(z) = \sum_r \psi_r z^{-r-1/2}
\end{equation}
satisfying
\begin{equation}
 \{\psi_m,\psi_n\} = \delta_{m+n,0}
\;.
\end{equation}
The Ramond algebra has a zero mode $\psi_0$ satisfying
$(\psi_0)^2=\frac 12$ and so there are two inequivalent irreducible
highest weight representations of the Ramond algebra, with highest
weights $\vec{\pm}$ satisfying
\begin{equation}
  \psi_0 \vec{\pm} 
= \pm \tfrac 1{\sqrt 2} \vec{\pm}
\;,\;\;
  \psi_m \vec{\pm} 
= 0 \hbox{ for $m>0$}.
\end{equation}
A unitary highest weight representation of the Ramond
algebra will have a highest weight space on which $\psi_0$ is
represented by a matrix with eigenvalues $\pm1$. We shall denote
the (vector-valued) highest weight of such a general representation by
$\vec\sigma$ and the chiral fields corresponding to such a state by
$\sigma(z)$. 

The energy-momentum tensor can be written in terms of $\psi$ as
\begin{equation}
 T(z) = \tfrac 12 \psi'(z) \psi(z)
\;,\;\;
\end{equation}
so that
\begin{equation}
 L_{-2}\vec 0 = \tfrac 12 \psi_{-3/2} \psi_{-1/2} \vec 0
\;,\;\;\;\;
 L_{-1} \vec{\pm} 
= \tfrac 1{2} \psi_{-1}\psi_0 \vec{\pm}
= \pm \tfrac 1{2\sqrt 2} \psi_{-1}\vec{\pm}
\label{ffsv}
\end{equation}
This last equation can be viewed as a null-vector equation, the
analogue of \eref{eq.nv12} and we can use it to find a differential
equation for the correlation functions of the form
\begin{equation}
F_\pm(z)= \cev\sigma \sigma(1) \sigma(z) \vec\pm
\;.
\end{equation}
Here, each $\sigma$ is some vector-value representation as is the
function $F$. We will not actually need to specify the exact form of these
representations, as we will see shortly.

We know that the space of Virasoro conformal blocks with the correct
properties is two dimensional so we have to consider $F_\pm(z)$ as a
vector in some space of solutions and the differential equation we
obtain will be a matrix differential equation.

According to the idea outlined above, we want to find
combinations of modes of the Ramond algebra $\Psi_\infty, \Psi_1$ and
$\Psi_z$ so that
\begin{eqnarray}
    \cev\sigma \, \sigma(1) \, \sigma(z) \, \Psi_\infty
&=& \cev\sigma\psi_0 \, \sigma(1) \, \sigma(z)
\;,
\\
    \cev\sigma \, \sigma(1) \, \sigma(z) \, \Psi_1
&=& \cev\sigma \, (\hat \psi_0 \sigma)(1)\,  \sigma(z)
\;,
\\
    \cev\sigma \, \sigma(1) \, \sigma(z) \, \Psi_z
&=& \cev\sigma \, \sigma(1) \, (\hat \psi_0 \sigma)(z)
\;.
\end{eqnarray}
We will require that these operators  square to $ 1/2$ and mutually
anti-commute, that is satisfy the algebra
\begin{equation}
 \{\Psi_\alpha,\Psi_\beta\} = \delta_{\alpha\beta}
\;.
\label{eq:psialg}
\end{equation}
We start by defining operators $\Phi_\alpha$ as integrals,
\begin{equation}
 \Phi_\alpha = \oint_0 p_\alpha(w)\psi(w)\D{w}
\;,
\end{equation}
where the contour encloses the origin but not the points $1$ or $z$ and
$p_\alpha(w)$ are functions which remove the unwanted singularities
in $w$ from the state
\begin{equation}
\cev\sigma \, \sigma(1) \, \sigma(z) \, \psi(w)
\;.
\end{equation}
Since the operator product of the field $\psi$ with a Ramond field
$\sigma$ is of the form
\begin{equation}
     \psi(z)\,\sigma(w)
\sim \frac{1}{\sqrt{z-w}}\hat\psi_0\sigma(w) + O(\sqrt{z-w})
\;,
\end{equation}
suitable combinations are
\begin{eqnarray}
    \Phi_\infty
&=& \oint_0 (1-w)^{1/2} (z-w)^{1/2} w^{-3/2} \psi(w) \D{w}
\nonumber\\
&=& \sqrt{z}\left(
    \psi_{-1} 
  - \frac{1+z}{2z} \psi_0
  - \frac{(1-z)^2}{8z^2} \psi_1 + \ldots \right)
\\
    \Phi_1
&=& \oint_0 (1-w)^{-1/2} (z-w)^{1/2} w^{-3/2} \psi(w) \D{w}
\nonumber\\
&=& \sqrt{z}\left(
    \psi_{-1} 
  - \frac{1-z}{2z} \psi_0
  - \frac{(1-z)(1+3z)}{8z^2} \psi_1 + \ldots \right)
\\
    \Phi_z
&=& \oint_0 (1-w)^{1/2} (z-w)^{-1/2} w^{-3/2} \psi(w) \D{w}
\nonumber\\
&=& \frac{1}{\sqrt{z}}\left(
    \psi_{-1} 
  + \frac{1-z}{2z} \psi_0
  + \frac{(1-z)(3+z)}{8z^2} \psi_1 + \ldots \right)
\end{eqnarray}
It is easy to calculate the anti-commutators of these operators --
either as contour integrals or directly in terms of the modes -- to
find
\begin{equation}
   (\Phi_\infty)^2 
= \frac 12 \;,\;\;
   (\Phi_1)^2
= -\frac{1-z}2 \;,\;\;
   (\Phi_z)^2
= \frac{1-z}{2z^3} \;,
\label{eq:phisq}
\end{equation}\begin{equation}
  \{ \Phi_\alpha,\Phi_\beta \}
= 0 \;, \;\;\;\alpha \neq \beta \;.
\end{equation}
Consequently we can define new combinations $\Psi_\alpha$ which
satisfy \eref{eq:psialg} as 
\begin{equation}
 \Psi_\infty = \Phi_\infty
\;,\;\;
 \Psi_1 = i\frac{1}{\sqrt{1-z}}\Phi_1
\;,\;\;
 \Psi_z = \frac{z^{3/2}}{\sqrt{1-z}}\Phi_z
\;.
\end{equation}

We can now use the combinations $\Psi_\alpha$ to replace the mode
$\psi_{-1}$ in the singular vector \eref{ffsv}.
There remains a large degree of choice in how to do this as we can
replace $\psi_{-1}$ by any of the $\Psi_\alpha$ as follows:
\begin{eqnarray}
    \psi_{-1}\vec\pm
&=& \left( \frac{1}{\sqrt z}\Psi_{\infty} + \frac{1+z}{2z}\psi_0
  \right)\vec\pm
\nonumber\\
&=& \left( -i\sqrt{\frac{1-z}{z}}\Psi_{1} + \frac{1-z}{2z}\psi_0
  \right)\vec\pm
\nonumber\\
&=& \left( \frac{\sqrt{1-z}}{z}\Psi_{z} - \frac{1-z}{2z}\psi_0
  \right)\vec\pm
\label{eq:subs}
\end{eqnarray}
Without loss of
generality, we will now just consider the correlation functions $F_+(z)$.
The general expression for the null vector relation 
\eref{ffsv} we can obtain in terms of $\Psi_\alpha$
using \eref{eq:subs} is of the form
\begin{equation}
 \left(
  L_{-1} 
  - \sum_{\alpha=1}^3 h_\alpha(z) \Psi_\alpha 
  - d(z)
 \right)\vec+ = 0
\;.
\label{ffsv2}
\end{equation}
where $h_\alpha(z)$ and $d(z)$ are functions of $z$.
Acting on this equation on the left by $\cev\sigma \sigma(1)\sigma(z)$
leads to a matrix differential equations for the correlation
functions of the form
\begin{equation}
  \left( 
  (z-1)\frac{\partial}{\partial z} + \frac 18 
- \sum_{\alpha=1}^3 \frac{1}{\sqrt 2}h_\alpha(z) \gamma_\alpha
- d(z)
  \right) F_+(z)
= 0
\;,  
\label{ffde}
\end{equation}
where $\gamma_\alpha/\sqrt 2$ are matrices
representing the action of the zero modes on the fields at $z$, 1 and
$\infty$ satisfying the Clifford algebra
\begin{equation}
\{\gamma_\alpha,\gamma_\beta\}=2 \delta_{\alpha\beta}
\;.
\label{eq:clifford}
\end{equation}
We must now choose a representation of this algebra. 
The smallest representations of this algebra are two dimensional and
there are two inequivalent choices for which $\gamma_\infty \gamma_1
\gamma_z = \pm i$.
It is essential for constructing
the correct differential equation for the correlation functions of the
spin field that the correct representation is chosen. 
To fix the equivalence class of the representation, we note that 
the operators $\Psi_\alpha$ satisfy further relations, for example 
\begin{equation}
  \left( \,
  \Psi_\infty \Psi_1 - i \Psi_z \psi_0 \,
  \right) \vec\sigma = 0
\;,\;\;
  \left( \,
  \Psi_\infty \Psi_1 \mp \tfrac{i}{\sqrt 2} \Psi_z \,
  \right) \vec\pm = 0
\;.
\end{equation}
Either of these is sufficient to fix the class of the representation.
Since we are considering $F_+(z)$, we note that
$\psi_0\vec+ = (1/\sqrt 2)\vec+$ implies we must choose the
representation of the Clifford algebra for which
$\gamma_\infty\gamma_1 = i \gamma_z$, ie for which 
$\gamma_\infty\gamma_1\gamma_z = i$.

Returning to equation \eref{ffde},
the simplest choices are for two of the functions $h_\alpha$ to be
zero, the other non-zero, so that only a single matrix
$\gamma_\alpha$ appears in the matrix differential equation
\eref{ffde}. In this case we can take the matrix to be diagonal 
(or alternatively consider the one-dimensional representations of the
algebra $(\Psi_\alpha)^2 = 1/2$) and we obtain a set of first order
differential equations for the correlation functions as follows which 
give the two components of
the function $F_+$ as the chiral blocks associated to a particular
channel. We illustrate this below.


\subsection{Using $\Psi_\infty$}
Setting $h_1=h_z=0$, we find $h_\infty=\sqrt{z/8}$ and 
$d=(1+z)/8z$, so that
the differential equation \eref{ffde} becomes
\begin{equation}
   \left( 
  (z-1)\frac{\partial}{\partial z} 
  -\frac 1{8z} - \frac1{4z}\gamma_\infty
  \right) F_+(z)
= 0
\;,  
\label{ffde1}
\end{equation}
At this stage the only requirement on $\gamma_\infty$ is that it
squares to $1$, so we can consider one-dimensional subspaces of the
space of correlation functions on which $\gamma_\infty$ takes values
$\pm 1$, leading to the two solutions
\begin{eqnarray}
\gamma_\infty=1
\;\;\;\;
&&
F_+ = \frac{\sqrt{1 - \sqrt z}}{z^{1/8}(1-z)^{1/8}} 
\\
\gamma_\infty=-1
\;\;\;\;
&&
F_+ = \frac{\sqrt{1 + \sqrt z}}{z^{1/8}(1-z)^{1/8}} 
\end{eqnarray}
which are the well--known chiral blocks of the Virasoro algebra
associated to the following choice of channel:
\[
\gamma_\infty=1 : \;\cbI{3500}{\infty}{1}{z}{0}{1/2}{}{}{}{}{}
\;,\qquad
\gamma_\infty=-1 : \;\cbI{3500}{\infty}{1}{z}{0}{0}{}{}{}{}{}
\;\;
\]


\subsection{Using $\Psi_1$}
The differential equation \eref{ffde} becomes
\begin{equation}
   \left( 
  (z-1)\frac{\partial}{\partial z} 
  +\frac {2z-1}{8z} - \frac i4 \sqrt{\frac{1-z}{z}}\gamma_1
  \right) F_+(z)
= 0
\;,  
\label{ffde2}
\end{equation}
Taking $\gamma_1$ diagonal with values
$\pm 1$ leads to the two solutions
\begin{eqnarray}
\gamma_1=1
\;\;\;\;
&&
F_+ = \frac{\sqrt{  \sqrt{z} +  \sqrt{z-1}  }}{z^{1/8}(1-z)^{1/8}}
\\
\gamma_1=-1
\;\;\;\;
&&
F_+ = \frac{\sqrt{  \sqrt{z} -  \sqrt{z-1}  }}{z^{1/8}(1-z)^{1/8}}
\end{eqnarray}
which are the chiral blocks of the Virasoro algebra
associated to the following channel:
\[
\gamma_1 = 1:\quad
\cbb{2300}{\infty}{z}{0}{1}{0}{}{}{}{}
\;,\qquad
\gamma_1 = -1:\quad
\cbb{2300}{\infty}{z}{1/2}{1}{0}{}{}{}{}
\]


\subsection{Using $\Psi_z$}
The differential equation \eref{ffde} becomes
\begin{equation}
   \left( 
  (z-1)\frac{\partial}{\partial z} 
  +\frac 1{8z} - \frac{\sqrt{1-z}}{4z}\gamma_z
  \right) F_+(z)
= 0
\;,  
\label{ffde3}
\end{equation}
Taking $\gamma_z$ diagonal with values
$\pm 1$ leads to the two solutions
\begin{eqnarray}
\gamma_z=1
\;\;\;\;
&&
F_+ = \frac{\sqrt{ 1 +  \sqrt{1-z}  }}{z^{1/8}(1-z)^{1/8}}
\\
\gamma_z=-1
\;\;\;\;
&&
F_+ = \frac{\sqrt{1 - \sqrt {1-z}  }}{z^{1/8}(1-z)^{1/8}} 
\end{eqnarray}
which are the well-known chiral blocks of the
Virasoro algebra associated to the channel:
\[
\gamma_z = 1:\quad
\cbb{2300}{1}{\infty}{0}{z}{0}{}{}{}{}
\;\;,\qquad
\gamma_z = -1:\quad
\cbb{2300}{1}{\infty}{1/2}{z}{0}{}{}{}{}
\]

\section{The Ramond chiral blocks}

Taking inspiration from the case of the Ising model, we will try to
find differential equations for the chiral blocks
\begin{equation}
F(z) = \cev{\lambda_\infty} \phi_{\lambda_1}(1) \phi_{\lambda_z}(z)
\vec{\lambda_{12}}
\;,
\label{eq:rcorr}
\end{equation}
The expectation is that we can express the singular vector
\eref{eq.nv12} in terms of
combinations of the modes $G_m$ which (inside the correlation
function) lead to a matrix representation of the algebra of the zero
modes acting on the primary fields inserted at $1$, $z$ and $\infty$.
That is, we will try to find combinations $\Gamma_\alpha$ of the
modes $G_m$ (for $\alpha\in\{\infty,1,z\}$)  which, when taken inside a
four-point function \eref{eq:rcorr} lead to a matrix representation $g_\alpha$ of the
zero-mode algebra 
\begin{equation}
\{g_\alpha,g_\beta\} 
= (h_\alpha - \frac c{24})\, \delta_{\alpha\beta}
= \lambda_\alpha^2\delta_{\alpha\beta}
\;,
\label{eq:gg}
\end{equation}
and which will lead to a matrix differential equation
for \eref{eq:rcorr} in the form
\begin{equation}
\left( \frac{\d}{\d z} + p(z) + \sum_\alpha q_\alpha(z)g_\alpha
\right) F(z) = 0
\;.
\end{equation}

The first step in repeating the analysis of the Ising model for the
superconformal algebra is to identify analogues of the combinations
$\Psi_\alpha$. From \eref{eq.hw.g.r}, 
 the operator product
of $G(z)$ with a Ramond field $\phi_\lambda(w)$ 
takes the form
\begin{equation}
G(z) \, \phi_\lambda(w)
=
  \frac{\hat G_0\phi_\lambda(w) }{(z-w)^{3/2}}
+ \frac{\hat G_{-1}\phi_\lambda(w) }{(z-w)^{1/2}}
+ O(\sqrt{z-w})
\;,
\label{eq:svirope}
\end{equation}
where $\hat G_0$ is a matrix representation of the zero mode
satisfying $\hat G_0^2 = \lambda^2 = h - c/24$, $\{\hat G_0,\hat
G_{-1}\}=0$ and $\hat G_{-1}^2 = L_{-2}$.
Furthermore,
\begin{equation}
  \cev\lambda G(z) 
= \cev\lambda \left( z^{-3/2}G_0 + z^{-1/2}G_1 + \ldots \right)
\end{equation}
Consequently, we are motivated to consider the three combinations
which remove all the singularities at two of the points $\infty, z$
and 1, and turn the leading singularity at the remaining point into a
simple pole:
\begin{eqnarray}
    \Psi_\infty
&=& \oint_0 (1-w)^{3/2}(z-w)^{3/2}w^{-5/2}G(w)\D w
\;,\nonumber\\
    \Psi_1 
&=& \oint_0 (1-w)^{1/2}(z-w)^{3/2}w^{-5/2}G(w)\D w
\;,\nonumber\\
    \Psi_z
&=& \oint_0 (1-w)^{3/2}(z-w)^{1/2}w^{-5/2}G(w)\D w
\;.
\end{eqnarray}
These operators, however, do not square to constants as in
\eref{eq:phisq}, nor do they simply anti-commute. Their algebra is more
complicated, and only simplifies inside the four-point functions \eref{eq:rcorr}. 
We can express their algebra in terms of suitable combinations of
modes of the Virasoro algebra:
\begin{eqnarray}
    e_\infty 
&=& \oint_0 \frac{ (1-w)^2 (z-w)^2 }{ w^3 } T(w) \D w
\;,\nonumber\\
    e_1 
&=& \frac{1}{(1-z)^2}\oint_0 \frac{ (1-w) (z-w)^2 }{ w^3 } T(w) \D w
\;,\nonumber\\
    e_z
&=& \frac{z^3}{(1-z)^2}\oint_0 \frac{(1-w)^2 (z-w) }{ w^3 } T(w) \D w
\;,\nonumber\\
    l_m
&=& \oint_0 \frac{ (1-w)^2 (z-w)^2 }{ w^{3+m} } T(w) \D w
\;,\;\; m\geq 1
\;.
\label{eq:esandls}
\end{eqnarray}
These combinations have the following properties when acting on the
state
\begin{equation}
 \cev\chi =  \cev{h_\infty}\, \phi_{h_1}(1) \,\phi_{h_z}(z)
\;,
\end{equation}
\begin{equation}
 \cev\chi (e_\infty - h_\infty)
=
 \cev\chi (e_1 - h_1)
=
 \cev\chi (e_z - h_z)
= 0
\;,\;\;
\end{equation}
\begin{equation}
\cev\chi l_m = 0\;,\;\; m\geq 1
\end{equation}
In terms of these operators, the $\Psi_\alpha$ satisfy
\begin{eqnarray}
    \Psi_\infty^2 
&=& (e_\infty - \frac{c}{24}) - (1+z)l_2 + z l_1
\;,\nonumber\\
    \Psi_1^2 
&=& -(1-z)^3 (e_1 - \frac{c}{24}) + zl_2 - (1-z) l_1
\;,\nonumber\\
    \Psi_z^2 
&=& \frac{(1-z)^3}{z^5}(e_z - \frac{c}{24}) +\frac 1z l_2 + \frac{1-z}{z^2} l_1
\;.
\end{eqnarray}
We thus define the combinations $\Gamma_\alpha$ as
\begin{equation}
\Gamma_\infty = \Psi_\infty
\;,\;\;
\Gamma_1 = -i(1-z)^{-3/2}\Psi_1
\;,\;\;
\Gamma_z = \frac{z^{5/2}}{(1-z)^{3/2}}\Psi_z
\;.
\label{eq:Gammas}
\end{equation}
Inside the four-point function \eref{eq:rcorr}, the terms in $l_2$ and
$l_{1}$ vanish 
and $(e_\alpha - \frac c{24}) = \lambda_\alpha^2$, so that
the action of the operators
$\Gamma_\alpha$ inside \eref{eq:rcorr} is given by matrices $g_\alpha$
satisfying the algebra 
\eref{eq:gg}.
To use these, we have to express the singular vector in terms of the
$\Psi_\alpha$. If we act with any of the $\Psi_\alpha$ in the highest
weight state $\vec{\lambda_{12}}$, the leading term is a multiple of
$G_{-3/2}\vec{\lambda_{12}}$, so that it is not possible to express
the singular vector in terms of just one of the
$\Psi_\alpha$. Instead, it is necessary to use all three and one finds
that
\begin{equation}
\textstyle
 G_{-1} \vec{\lambda_{12}}
 = \left(
 \frac{1+z}{2z} \lambda_{12}
 - i \left(\frac{1-z}z\right)^{1/2} \Gamma_1
 - \frac{(1-z)^{1/2}}z \Gamma_z
 + z^{-1/2} \Gamma_\infty
   \right)\vec{\lambda_{12}}
\;.
\label{eq:g-1}
\end{equation}
Combining \eref{eq:g-1} with \eref{eq:l-1}
leads to the following matrix differential equation for
\eref{eq:rcorr}, the main result of this article:
\begin{equation}
{
\begin{array}{rcl}
&&\\[-3mm]
&& (z-1) F' + 
\textstyle
(h_{12} + h_1 + h_z - h_{\infty}) F(z)
\nonumber \\[2mm]
 &+&
\textstyle
\sqrt{\frac t2}\left[
 \frac{1+z}{2z} \lambda_{12}
 - i \left(\frac{1-z}z\right)^{1/2} \!\!\!\! \!\!g_1
 - \frac{(1-z)^{1/2}}z g_z
 + z^{-1/2} g_\infty
   \right]
  F(z) = 0
\;. \\[3mm]
\end{array}
}
\label{eq:gde}
\end{equation}
We now turn to the analysis of this equation and its solutions.
It will also be convenient to parametrise $\lambda_\alpha$ 
according to \eref{eq:lpq} as 
\begin{equation}
\lambda_\infty = \lambda_{1,q}
\;,\;\;
\lambda_1 = \lambda_{1,r}
\;,\;\; \hbox{ and }
\lambda_z = \lambda_{1,s}
\;.
\label{eq:lamparam}
\end{equation}
The first thing we can do is to check the indices of the solutions
around the points $\infty,1$ and $0$, and then we can find the
explicit solutions in various cases. Finally we compare these to
solutions known by other methods. 

\subsection{Indicial equations}

These are the equations for the leading behaviour of the solution
around a singular point. The equation \eref{eq:gde} has singular
points $0, 1$ and $\infty$. We first consider the point $0$ around
which the solution will have an expansion of the form
\begin{equation}
F(z) = \sum_{n=0}^\infty a_n z^{\alpha + n/2}
\;,
\label{eq:Fexp}
\end{equation}
where $a_n$ are vectors. Note that the expansion will be in
half-integer powers of $z$ as the differential equation explicitly
contains $\sqrt z$.
Around $z=0$, the leading behaviour
is determined by substituting \eref{eq:Fexp} in  \eref{eq:gde} and examining
the coefficient of $z^{\alpha-1}$:
\begin{eqnarray}
( - \alpha  + 
\textstyle
\sqrt{\frac t2}\left[
 \frac{1}{2} \lambda_{12}
 - g_z
   \right]
 ) a_0 = 0
\;.
\label{eq:gde2} 
\end{eqnarray}
Since this equation only involves $g_z$, we can take $g_z$ to be
diagonal with eigenvalues $\pm\lambda_z$. This leads to two
solutions for $\alpha$,

\begin{equation}
\alpha_\pm = 
\textstyle
\sqrt{\frac t2}\left[
 \frac{1}{2} \lambda_{12}
 \mp \lambda_z
   \right]
\;.
\end{equation}
If we use the parametrisation $\lambda_z = \lambda_{1,s}$, we find
\begin{eqnarray}
\alpha_+ = \frac 18 (2t(s-1) - 1)
&& = h_{1,s+1} - h_{1,s} - h_{1,2}
\;,
\\
\alpha_- = \frac 18 (3 - 2t(s+1))
&& = h_{1,s-1} - h_{1,s} - h_{1,2}
\;,
\end{eqnarray}
These values are exactly the expected exponents for the chiral blocks
shown below:

\vskip 5mm
\begin{equation}
\alpha_+:\quad
\cbb{2500}{\lambda_{1,q}}{\lambda_{1,r}}{h_{1,s+1}^+}{\lambda_{1,s}}{\lambda_{1,2}}{\epsilon}{+}
\qquad,\;\;
\alpha_-:\quad
\cbb{2500}{\lambda_{1,q}}{\lambda_{1,r}}{h_{1,s-1}^+}{\lambda_{1,s}}{\lambda_{1,2}}{\epsilon}{-}
\end{equation}
It is easy to check that similar results hold for the other two
channels, corresponding to expanding the solution $F$ around $\infty$
in powers of $1/z$ and around $1$ in powers of $(z-1)$.

\subsection{Solution to the matrix differential equation}

To solve the full equation \eref{eq:gde} we must choose a matrix
representation for $g_\alpha$. 
Up to now we have not had to specify the action of the zero modes on
the primary fields, and as in the Ising model, we do not need to do it
now. There are only two
inequivalent representations for which $g_\infty g_1 g_z = \pm i
\lambda_\infty \lambda_1 \lambda_z$, and the choice of representation is 
invariant under monodromy around $z=0$ and $z=1$.
We shall take $g_\alpha$ to be given in terms of the Pauli matrices as
\begin{eqnarray}
   g_\infty 
&=& \lambda_\infty \sigma^1 
= \lambda_\infty \pmatrix{0 & 1 \cr 1 & 0}
\;,\;\;
\nonumber\\
   g_1
&=& \ggg \lambda_1 \sigma^2
= \ggg \lambda_1 \pmatrix{0 & -i \cr i & 0}
\;,\;\;
\nonumber\\
   g_z
&=& \lambda_z \sigma^3
= \lambda_z \pmatrix{1 & 0 \cr 0 & -1}
\;,
\label{eq:gdefs}
\end{eqnarray}
where $\ggg=\pm 1$. This choice turns the equation for $F$ into a
real matrix differential equation and 
taking $g_z$ diagonal also means that the two components of $F$ have
expansions in $z$ (and not $\sqrt z$). The two components of $F$ can
be identified by comparison with \eref{eq:ff} and \eref{eq:GG} and are the 
odd and even chiral blocks (up to a sign) 
\begin{eqnarray}
\alpha=\alpha_+
&:&
F = \pmatrix{
{\mbox{
\cbb{2700}{\lambda_{1q}}{\lambda_{1r}}{h_{1,s{+}1}^\Even}{\lambda_{1s}}{\lambda_{12}}{\ggg}{+}
}}
\cr 
{\mbox{
\cbb{2700}{\lambda_{1q}}{\lambda_{1r}}{h_{1,s{+}1}^\Odd}{\lambda_{1s}}{\lambda_{12}}{\ggg}{+}
}}
}
\\
\alpha=\alpha_-&:&
F = \pmatrix{
-\ggg\;\; \hbox to 0pt{$\cdot$}& 
{\mbox{
\cbb{2700}{\lambda_{1q}}{\lambda_{1r}}{h_{1,s{-}1}^\Odd}{\lambda_{1s}}{\lambda_{12}}{-\ggg}{-}
}}
\cr 
& {\mbox{
\cbb{2700}{\lambda_{1q}}{\lambda_{1r}}{h_{1,s{-}1}^\Even}{\lambda_{1s}}{\lambda_{12}}{-\ggg}{-}
}}
}
\end{eqnarray}

\section{Exact solutions}

In some cases it is possible to find exact solutions to
\eref{eq:gde}. 
The simplest case to consider is where all the $\lambda_\alpha$ are
equal to $\lambda_{1,2}$.
In this case
\begin{equation}
h_{1,2} = \frac{3}{16}(2t-1)
\;,\;\;
\lambda_{1,2} = \frac{1-2t}{2\sqrt{2t}}
\;,\;\;
\sqrt{\tfrac t2}\lambda_{1,2} = -\frac 43 h_{1,2}
\;,\;\;
\end{equation}
and the fusion rules
\eref{eq:+channel}
force $\ggg=+$.
Writing the components of $F$
as $(f_1,f_2)$, with the representation \eref{eq:gdefs}, the
differential equation \eref{eq:gde} becomes 
\begin{equation}
   (z-1) \pmatrix{f_1' \cr f_2'} 
  + 2h \pmatrix{f_1 \cr f_2}
  -\frac{ 4h}3
\pmatrix{
  \frac{1+z}{2z} 
{-}
  \frac{\sqrt{1-z}}z 
&
  \frac{1}{\sqrt z} 
{-}
  \frac{\sqrt{1-z}}{\sqrt z}
 \cr
  \frac{1}{\sqrt z} 
{+}
  \frac{\sqrt{1-z}}{\sqrt z}
&
 \frac{1+z}{2z} 
{+}
  \frac{\sqrt{1-z}}z 
}
\pmatrix{f_1 \cr f_2} = \pmatrix{0 \cr 0}
\label{eq:mateq}
\end{equation}
The coefficients in
the equation as presented have branch cuts at $z=0$ and $z=1$, and it
is convenient to remove these by changing variables to
\begin{equation}
 u = \frac{1 - \sqrt{1-z}}{z} = \frac{\sqrt z}{2} + \ldots
\;,\;\;\;\;
 z = \frac{4 u^2}{(1 + u^2)^2}
\;.
\end{equation}
With this change and a redefinition $F(z) = z^{-2h} G(u)$ 
to remove
the leading singularity in $z$, the equations simplify dramatically to
\begin{eqnarray}
-\frac{(1 - u^4)}{8 u}\,\frac{\d G}{\d u}
+ \frac{2h}{3}
  \pmatrix{ {1}/{u^2} & -{2u} \cr
            -{2}/{u} & u^2 } G = 0 
\end{eqnarray}
with solutions
\newcommand{\FFF}{ {}_2F_1 }
\renewcommand{\FFF}{ F\! }
\begin{eqnarray}
G^+ = 
u^{-1+2 t} \left(1{-}u^4\right)^{\frac{3}{4}{-}\frac{3 t}{2}}
\pmatrix{
\FFF\left(\frac{1}{2}{-}t,\frac{1}{2}{-}\frac{t}{2};\frac{t}{2};u^4\right)
\cr
\frac{1{-}2t}{t}\cdot
u  \cdot
\FFF
\left(\frac{3}{2}{-}t,\frac{1}{2}{-}\frac{t}{2};1{+}\frac{t}{2};u^4\right)
}
\\[3mm]
G^- = \left(1{-}u^4\right)^{\frac{3}{4}-\frac{3 t}{2}}
\pmatrix{
\frac{2t{-}1}{t{-}2} \cdot
u^3 \cdot 
\FFF\left(\frac{3}{2}{-}t,\frac{3}{2}{-}\frac{3t}{2};2{-}\frac{t}{2};u^4\right)
\cr
\FFF\left(\frac{1}{2}{-}t,\frac{3}{2}{-}\frac{3t}{2};1{-}\frac{t}{2};u^4\right)
}
\end{eqnarray}
where $\FFF$ is the standard Hypergeometric function.
For special values of $t$, the blocks with identity intermediate
channel simplify further. For example at $t=5/3$, that is for
four-point blocks of the field with $h=7/16$ in the tri-critical
Ising model, $G^-$ simplifies to
\begin{equation} 
G^- = 
{\left(1-u^4\right)^{-7/4}}
\pmatrix{
-{u^3 \left(7+u^4\right)}
\cr
 {(1+7 u^4)}
}
\end{equation}
As can be seen, the two components are related by $u\to1/u$ which
corresponds to $z$ encircling the branch point at $z=1$ once. 

The blocks can also be found exactly in the more general case
$\lambda_\infty = \lambda_1$, that is $q=r$ in the parametrisation
\eref{eq:lamparam}. 
For example, the solutions for $F$ with $\ggg=+$ are
\begin{eqnarray}
F^+ &=& 
u^{-\frac{1}{4}{+}\frac{1}{2} (s{-}1) t} \left(1{-}u^2\right)^{\frac{1}{4}
  \left(2 s{-}1-2 (r{-}1) t-s^2 t\right)}
\left(1{+}u^2\right)^{\frac{1}{4} (4-2 r t{+}s (s t{-}2))}
\\
&&
\pmatrix{
\FFF\left(
  \frac{1{-}rt}{2},\frac{2 {-} 2rt {+} st}{4};
  \frac{st}{4};u^4\right)
\cr
\frac{2(1 {-} rt)}{st}\cdot
u  \cdot
\FFF\left(
  \frac{3{-}rt}{2},\frac{2 {-} 2rt {+} st}{4};
  \frac{4 {+} st}{4};u^4\right)
}
\\[3mm]
F^- &=& 
u^{\frac{3}{4}{-}\frac{1}{2} (s{+}1) t} 
\left(1{-}u^2\right)^{\frac{1}{4} \left(2 s{-}1{-}2 (r{-}1) t{-}s^2 t\right)}
\left(1{+}u^2\right)^{\frac{1}{4} (4{-}2 r t{+}s (s t{-}2))}
\\
&&
\pmatrix{
\frac{2rt{-}2}{st{-}4} \cdot
u^3 \cdot 
\FFF
\left(\frac{3{-}rt}{2},\frac{6{-}2rt{-}st}{2};\frac{8{-}st}{4};u^4\right)
\cr
\FFF\left(\frac{1{-}rt}{2},\frac{6{-}2rt{-}st}{2};\frac{4{-}st}{4};u^4\right)
}
\end{eqnarray}

\section{Solutions known by other methods}

We can check the differential equation and its solutions against 
solutions known by other methods. Principally, there are three values
of the central charge 
in the superconformal minimal series which also appear in list of
Virasoro minimal models, so that A-series of the superconformal
minimal models can be identified with the following invariants of the
Virasoro minimal models,
\begin{eqnarray}
&&
SM(3,5) = M(4,5)_D
\;,\;\;
\nonumber\\
&&
SM(2,8) = M(3,8)_D
\;,\;\;
\nonumber\\
&&
SM(3,7) = M(7,12)_E
\;.
\end{eqnarray}
The representations in the superconformal models can be found as sums
of representations in the Virasoro minimal models, and the chiral blocks of the
superconformal models must be sums of Virasoro chiral blocks. Power
series expansions of the Virasoro chiral blocks can be found easily,
either by using one of the recursion relations of Zamolodchikov \cite{Zam1}
or by solving the differential equation from the singular vector. 
We present two examples here to show how this works.

\subsection{$SM(3,5)$}

In this model, the tricritical Ising model, the Ramond representations
of the superconformal algebra are
$(1,2)$ and $(2,1){\equiv}(1,4)$ with conformal dimensions $h_{1,2}=3/80$ and
$h_{1,4}=7/16$, and consequently all correlation functions of four
Ramond fields can be found using the method in this paper. 

If we consider just one case, the following two blocks can be found by
series solution of the differential equation \eref{eq:mateq}
\begin{equation}
\cbb{2300}{\frac 3{80}}{\frac 7{16}}{\frac{1}{10}^\Even}{\frac 7{16}}{\frac 3{80}}{-}{-}
\quad
 = 
z^{-3/8} \big(
1+\frac{5 z}{4}+\frac{75 z^2}{64}+\frac{287 z^3}{256}+
\frac{8885 z^4}{8192} + \ldots
\big)
\label{eq:block1}
\end{equation}
\begin{equation}
\cbb{2300}{\frac 3{80}}{\frac 7{16}}{\frac{1}{10}^\Odd}{\frac 7{16}}{\frac 3{80}}{-}{-}
\quad
 =
z^{1/8} \big(
1+\frac{5 z}{6}+\frac{149 z^2}{192}+\frac{95 z^3}{128}+\frac{5885 z^4}{8192}
+ \ldots
\big)
\label{eq:block2}
\end{equation}
Since the Ramond representations of the unextended superconformal
algebra and the  even and odd sectors of the intermediate channel are
each irreducible representations of the Virasoro algebra,
these blocks can also be found using the representation theory of the Virasoro
algebra. In this case, the irreducible Virasoro representation of
weight $3/80$ has Virasoro Kac labels 
$(2,2)_{V}$ and so the odd and even chiral blocks are two solutions of 
a fourth order differential equation. This differential equation can
also be solved for a  
series solution and two of the solutions are exactly those given in 
\eref{eq:block1} and \eref{eq:block2}.

\subsection{$SM(3,7)$}

This model is related to the $E_6$ invariant of the Virasoro minimal
model $M(7,12)$. 
The irreducible
super Virasoro representations of interest split into direct sums of
irreducible Virasoro representations as follows
\begin{equation}
\begin{array}{rlrl}
\cH_{(11)}^\Even =& \cH^{Vir}_{(11)} \oplus \cH^{Vir}_{(17)}
\;,\;\;&
\cH_{(11)}^\Odd =& \cH^{Vir}_{(15)} \oplus \cH^{Vir}_{(1,11)}
\;,\;\;\\
\cH_{(12)} =& \cH^{Vir}_{(24)} \oplus \cH^{Vir}_{(28)}
\;,\;\;\\
\cH_{(13)}^\Even =& \cH^{Vir}_{(35)} \oplus \cH^{Vir}_{(3,11)}
\;,\;\;&
\cH_{(13)}^\Odd =& \cH^{Vir}_{(31)} \oplus \cH^{Vir}_{(37)}
\;,\;\;\\
\cH_{(14)} =& \cH^{Vir}_{(34)} \oplus \cH^{Vir}_{(38)}
\;,\;\;\\
\cH_{(15)}^\Even =& \cH^{Vir}_{(25)} \oplus \cH^{Vir}_{(2,11)}
\;,\;\;&
\cH_{(15)}^\Odd =& \cH^{Vir}_{(21)} \oplus \cH^{Vir}_{(27)}
\;,\;\;\\
\cH_{(16)} =& \cH^{Vir}_{(14)} \oplus \cH^{Vir}_{(18)}
\;,\;\;
\;.
\end{array}
\end{equation}
Most of the superconformal chiral blocks are given by
sums of Virasoro chiral blocks, but in some cases there is only a
single Virasoro representation contributing to the intermediate
channel and so the results of solving \eref{eq:mateq} and the Virasoro
null vector equations can be compared directly.
If we denote Virasoro representations by $(rs)_V$ then two such cases are 
\begin{eqnarray}
\cbb{2300}{\scriptstyle(12)}{\scriptstyle(14)}{\scriptstyle
  (13)^\Even}{\scriptstyle(12)}{\scriptstyle(12)}{+}{-}
\quad
&=&
\quad
\cbb{2300}{\scriptstyle(24)_V}{\scriptstyle(34)_V}{\scriptstyle
  (35)_V}{\scriptstyle(24)_V}{\scriptstyle(24)_V}{}{}
\nonumber\\
 &=& 
z^{-1/56} 
\big(
1+\frac{z}{28}+\frac{107 z^2}{3136}+
\frac{2523 z^3}{87808}
%
+ \ldots
\big)
%
\label{eq:block4}
\end{eqnarray}
\begin{eqnarray}
\cbb{2300}{\scriptstyle(12)}{\scriptstyle(14)}{\scriptstyle (15)^\Even}{\scriptstyle(14)}{\scriptstyle(12)}{+}{+}
\quad
&=&
\quad
\cbb{2300}{\scriptstyle(24)_V}{\scriptstyle(34)_V}{\scriptstyle
  (25)_V}{\scriptstyle(34)_V}{\scriptstyle(24)_V}{}{}
\nonumber\\
&=& 
z^{11/28} 
\big(
1+\frac{9 z}{28}+\frac{25309 z^2}{122304}+\frac{536597
 z^3}{3424512}
%
+ \ldots
\big)
%
\label{eq:block3}
\end{eqnarray}
This time the blocks 
\eref{eq:block4} and \eref{eq:block3}  
can be calculated either as series solutions
of \eref{eq:mateq} or as 
the series solutions of eighth order differential equations
corresponding to the level eight null vector in the Virasoro
representation $(2,4)_{V}$, giving the same answers shown.
They can also be compared with
the general series solution for Virasoro chiral blocks found by
Zamolodchikov \cite{Zam1}.

\section{Conclusions}

We have found a matrix differential equation for Ramond four-point
chiral blocks. This can solved completely in some cases but as yet the
full general solution is now known.
These solutions were known already exactly in some other cases and as
series solutions in a few more cases based on relations with Virasoro
minimal models. The differential equations presented here reproduce
these known results.
Exact integral formulae based on free-field constructions are also
known  \cite{MSSt2} and it remains to check that these satisfy the
equations we have 
found.

Recently, recursive formulae generalising
Zamolodchikov's elliptic recursion formulae in \cite{Zam1} have been
found \cite{Elliptic} and again
it remains to check that these give series expansions which satisfy
our equations.

For the future, these equations and their solutions should enable one
to extend calculations that have only fully been worked out in the
Virasoro case to the superconformal case, such as the construction of
the full set of boundary structure constants in \cite{Runk1,Runk2}

\ack

GMTW is very grateful to
F.~Wagner and 
I.~Runkel for
discussions on many aspects of fusion and superconformal field theory
and to IR for detailed comments on the manuscript.
This work was supported by PPARC and STFC through rolling grants 
PP/C507145/1 and ST/G000395/1 and by the studentship PPA/S/S/2005/4103.

\end{document}